\documentclass[apjl]{emulateapj}
\usepackage{amsmath}
\usepackage{amssymb}
\usepackage{bm}

\usepackage{color}
\usepackage[varg]{txfonts}
  
\begin{document}
\label{firstpage}

\title[Generalised warped disk equations]{Generalised warped disk equations}

\author{Rebecca G. Martin\altaffilmark{1,4}}
\author{Stephen H. Lubow\altaffilmark{2}}
\author{J. E. Pringle\altaffilmark{3,4}}
\author{Alessia Franchini\altaffilmark{1}}
\author{Zhaohuan Zhu\altaffilmark{1}}
\author{Stephen Lepp\altaffilmark{1}} 
\author{Rebecca Nealon\altaffilmark{4}}
\author{C. J. Nixon\altaffilmark{4}}
\author{David Vallet\altaffilmark{5}}
\affil{\altaffilmark{1}Department of Physics and Astronomy,
  University of Nevada, Las Vegas, 4505 South Maryland Parkway, Las
  Vegas, NV 89154, USA}
\affil{\altaffilmark{2}Space Telescope Science Institute, 3700
  San Martin Drive, Baltimore, MD 21218, USA}
\affil{\altaffilmark{3}Institute of
  Astronomy, Madingley Road, Cambridge, CB3 0HA}
\affil{\altaffilmark{4}Department of
  Physics and Astronomy, University of Leicester, University Road,
  Leicester LE1 7RH, UK}
\affil{\altaffilmark{5}Department of Mechanical Engineering,
  University of Nevada, Las Vegas, 4505 South Maryland Parkway, Las
  Vegas, NV 89154, USA}

\begin{abstract}
The manner in which warps in accretion disks evolve depends on the magnitude of the viscosity. For small viscosity $(\alpha < H/R)$ the warp evolves in a wave-like manner; for large viscosity $H/R<\alpha \ll 1$ it evolves diffusively. Here $\alpha$ is the viscosity parameter and $H/R$ the disk aspect ratio. Currently there is no simple set of equations which describes the evolution in both regimes. In this paper we describe a possible solution to this problem and introduce a set of one--dimensional equations that describe the evolution of a warped disk that are applicable in both high and low viscosity regimes for arbitrary tilts, but small warps.
\end{abstract}


\section{Introduction}
\label{intro}

Warped disks are expected to occur in a large number of astrophysical
situations \citep[e.g.][]{Pringle1981,Pringle1999,Kingetal2013}.
Warping may occur due to external torques from various
sources. Binaries can provide such a torque on a circumstellar disk
from an external binary component
\citep[e.g.][]{PT1995,Larwoodetal1996,LO2000,Martinetal2009,Martinetal2011} 
or a torque on a circumbinary disk form an internal binary
\citep[e.g.][]{Facchinietal2013,Lodato2013,Martin2017}. Around spinning black
holes, general relativistic Lense--Thirring precession may cause
warping \citep{BP1975} in X--ray binaries
\citep[e.g.][]{SF1996,WP1999, Ogilvie2001, Martinetal2007} and around
supermassive black holes \citep[e.g.][]{Herrnstein1996,Martin2008}.
Disks in AGN and in binary X-ray sources may be warped by the effects
of radiation pressure \citep{Pringle1996,OD2001}. Disks may also be
warped by a misaligned magnetic field \citep[e.g.,][]{Lai1999} or a planet \citep[e.g.][]{Nealon2018}.

The evolution of the warp in a disk depends upon how the \cite{SS1973}
viscosity $\alpha$ parameter compares with the disk aspect ratio
$H/R$. If $\alpha>H/R$, then the warp propagates diffusively through
the effects of viscosity.  If $\alpha<H/R$, then pressure forces drive
the evolution and the warp propagates as a bending wave that travels
at half the sound speed, $c_{\rm s}/2$ \citep{PL1995,Pringle1999,LO2002}. In this
case, the viscosity is too small to damp the wave locally. For a
recent review of warped disks, see \cite{Nixon2016}.

One-dimensional (based on radius) models of disk tilt evolution offer
advantages over multi-dimensional models. They permit tracking the
evolution over long timescales with much less computational effort
than multi-dimensional models. They are also easier to interpret
physically.  On the other hand, their applicability is limited by the
simplifications made to reduce the dimensionality. In any case, such
models can be compared with multi-dimensional models to obtain more
physical insight.

A one-dimensional model should ideally conserve angular momentum and
be valid for arbitrary tilts and warps (the derivative of the tilt
with respect to the logarithm of the radius).  In the viscous regime,
\cite{Pringle1992} developed a set of intuitively-based
one-dimensional equations that satisfy these conditions.
\cite{Ogilvie1999} extended this analysis by directly working with the
fluid equations and obtained one-dimensional equations that apply for
even large warps. This analysis showed that the \cite{Pringle1992}
equations are valid for small warps and small $\alpha$, $H/r < \alpha
\ll 1$, but arbitrary tilts, with the extension that the effective
viscosity coefficients are constrained by the internal fluid
dynamics.

In application to disks around young stars, the wave--like regime is
of importance.  One-dimensional linear disk evolution equations for
this regime typically assume that the warp is small and ignore disk
surface density evolution \citep[e.g.,][]{PL1995,
  LO2000}. \cite{Ogilvie2006} analyzed the nonlinear dynamics of free
warps (imposed by initial conditions) in the absence of viscous
density evolution.  However, as found in \cite{Bateetal2000},
significant density evolution can occur as the tilt evolves.
We are interested in the case that the disk is in good radial
communication so that the level of warping is small, as should apply
to protostellar disks. Therefore a linear analysis is often valid.  We
are interested in the case that the disk tilt and surface density
change over the course of its evolution.

The goal of this work is to find a formulation 
which describes the disk evolution correctly in both regimes, and manages to 
connect the two. In Section~\ref{equations} we present two sets of warped disk equations, one valid in the viscous regime  and one valid in the wave--like regime.
In Section~\ref{general}, we follow along the lines of \cite{Pringle1992} to extend the linear tilt evolution equations to apply to arbitrary tilts and account for viscous density evolution.  
 In Section~\ref{numerical} we
numerically solve the equations for an initially warped disk around a
single central object in the absence of any external torques. We draw
our conclusions in Section~\ref{conc}.

\section{Warped disk equations in the two regimes}
\label{equations}

Currently there are two sets of warped disk equations that describe
mutually exclusive regimes, the wave--like regime with $\alpha < H/R$
and the diffusive (or viscous) regime with $\alpha > H/R$. We provide
an overview of these two sets of equations in this Section.  We
describe the disk as consisting of a set circular rings with spherical
radius $R$. We assume that the disk is in near Keplerian
rotation. \footnote{ Thus our results do not apply to strongly
  non--Keplerian flows such as those that occur in accretion disks
  close to the event horizon of a black hole.}  The rings rotate with
Keplerian angular frequency $\Omega(R)=\sqrt{G M/R^3}$ about a central
object of mass $M$ and have a surface density $\Sigma(R)$.  The disk
extends from inner radius $R_{\rm in}$ to outer radius $R_{\rm
  out}$. The angular momentum per unit area of each ring is
\begin{equation}
\bm{L}=\Sigma R^2 \Omega \bm{l},
\label{eqL}
\end{equation}
where $\bm{l}$ is a unit vector. We consider a locally isothermal disk with aspect ratio $H/R$, where $H$ is the disk scale height.

Angular momentum transport in an accretion disk is driven by turbulent
eddies with a maximum size $H$, and maximum speed the sound speed,
$c_{\rm s}=H \Omega$. The azimuthal shear viscosity has the standard
form
\begin{equation}
    \nu_1 = \alpha_1 \left(\frac{H}{R}\right)^2 R^2 \Omega
\end{equation}
\citep{SS1973}, where $\alpha_1 \simeq \alpha$, for dimensionless
parameter $\alpha<1$.  The vertical shear viscosity is
\begin{equation}
    \nu_2 = \alpha_2 \left(\frac{H}{R}\right)^2 R^2 \Omega,
\end{equation} 
\citep{PP1983}, where $\alpha_2\simeq 1/(2\alpha)$ in the linear approximation.

\subsection{Wave--like limit equations}

In the wave--like limit, $\alpha <H/R$, we assume that the surface density does not evolve, $\partial \Sigma/ \partial t=0$.   The evolution of a warped disk is described by  two equations
\begin{align}
\frac{\partial \bm{G}}{\partial t}+ \omega \, \bm{l}\times \bm{G} + \alpha \Omega \bm{G}=\frac{\Sigma H^2 R^3\Omega^3}{4}\frac{\partial \bm{l}}{\partial R}
\label{lo1}
\end{align}
and
\begin{align}
\Sigma R^2 \Omega  \frac{\partial \bm{l} }{\partial t} =
\frac{1}{R}\frac{\partial \bm{G}}{\partial R}+\bm{T},
%+ \bm{T}_{\rm p},
\label{lo2}
\end{align}
where $\bm{G}$ is the internal disk torque and $\bm{T}$ is the external torque on the disk \citep[see equations~12 and~13 in][]{LO2000}.
  The apsidal precession frequency in the plane of the disk is
\begin{equation}
\omega =\frac{\Omega^2-\kappa^2}{2 \Omega}
\end{equation}
with epicyclic frequency $\kappa$. We solve equations~(\ref{lo1}) and~(\ref{lo2}) in Section~\ref{wl} to compare to our solution to the generalised equations in the wave--like limit.

\subsection{Viscous limit equations}

In the viscous limit, $\alpha > H/R$, a warped disk is described by the evolution equation
\begin{align}
  \frac{\partial \bm{L} }{\partial t} = &
  \frac{3}{R}\frac{\partial}{\partial R} \left[\frac{R^{1/2}}{\Sigma} \frac{\partial}{\partial R}  \left(\nu_1 \Sigma R^{1/2} \right) \bm{L}  \right]  \cr 
 & +\frac{1}{R}\frac{\partial}{\partial R} \left[
  \left( \nu_2 R^2 \left|\frac{\partial \bm{l}}{\partial R} \right|^2-\frac{3}{2}\nu_1 \right) \bm{L}  \right] \cr
& +\frac{1}{R}\frac{\partial}{\partial R} \left[
  \frac{1}{2}\nu_2 R |\bm{L}|\frac{\partial \bm{l}}{\partial R} \right] 
+\bm{T}
\label{viscous}
\end{align}
\citep{Pringle1992}. We solve equation~(\ref{viscous}) in Section~\ref{diff} to compare to our solution to the generalised equations in the viscous regime.

\section{Generalised warped disk equations}
\label{general}

We now show how it is possible to combine the above equations into
a single set that are valid in both the wave--like and the diffusive warp
propagation regimes.  We follow the methods of \cite{PP1983} and \cite{Pringle1992}. We note that this is not a first principles derivation of the
evolution equations \citep[cf.,][]{Ogilvie1999, Ogilvie2006}. Conservation of mass is expressed as
\begin{equation}
\frac{\partial \Sigma}{\partial t}+\frac{1}{R}\frac{\partial}{\partial R}(R\Sigma v_R)=0,
\label{mass}
\end{equation}
where $v_R$ is the radial velocity. Conservation of angular momentum gives us
\begin{equation}
\frac{\partial \bm{L}}{\partial t }+\frac{1}{R}\frac{\partial }{\partial R}(\Sigma v_R R^3 \Omega \bm{l})
=\frac{1}{R}\frac{\partial \bm{G}}{\partial R} +\bm{T},
\label{angmom}
\end{equation}
where $\bm{G}$ is the internal disk torque and $\bm{T}$ is the
external torque on the disk. In equation (\ref{angmom}), we have
included the second term on the LHS compared to equation~(\ref{lo2}), in the wave--like equations,  in order to enforce conservation of angular momentum as the disk density evolves.

\begin{figure*}
\centering
\includegraphics[width=7.5cm]{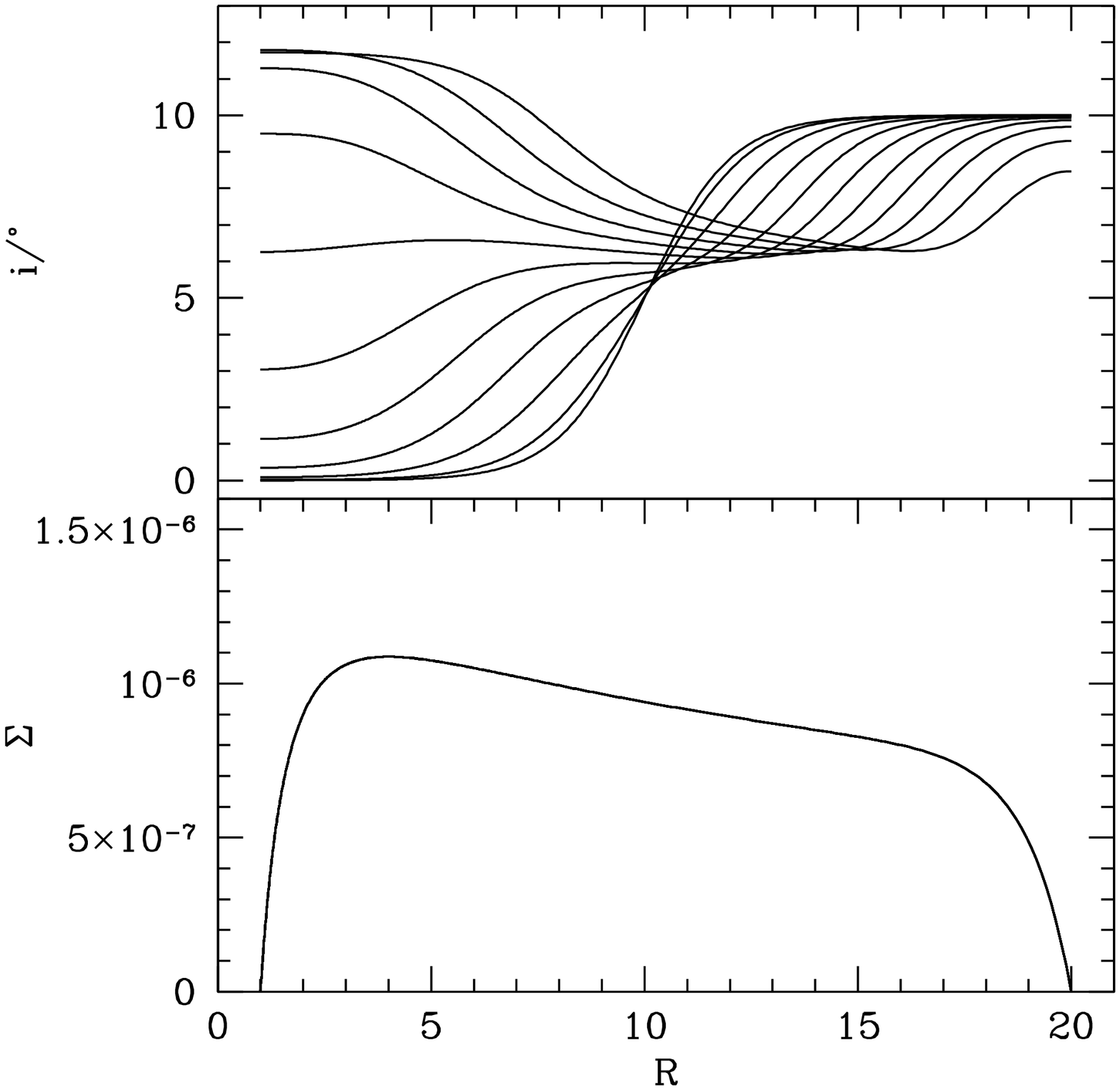}
\includegraphics[width=7.5cm]{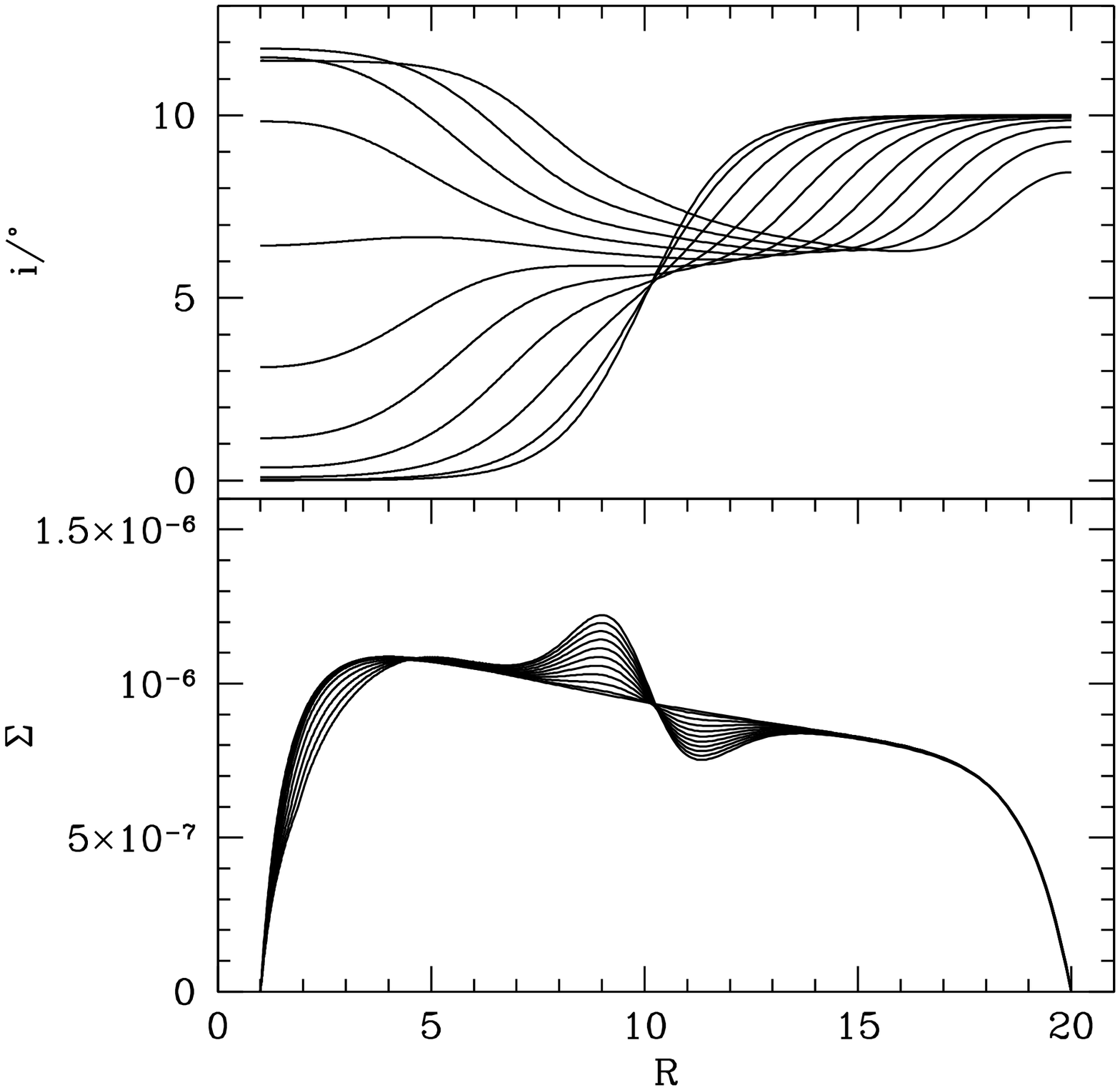}
\includegraphics[width=7.5cm]{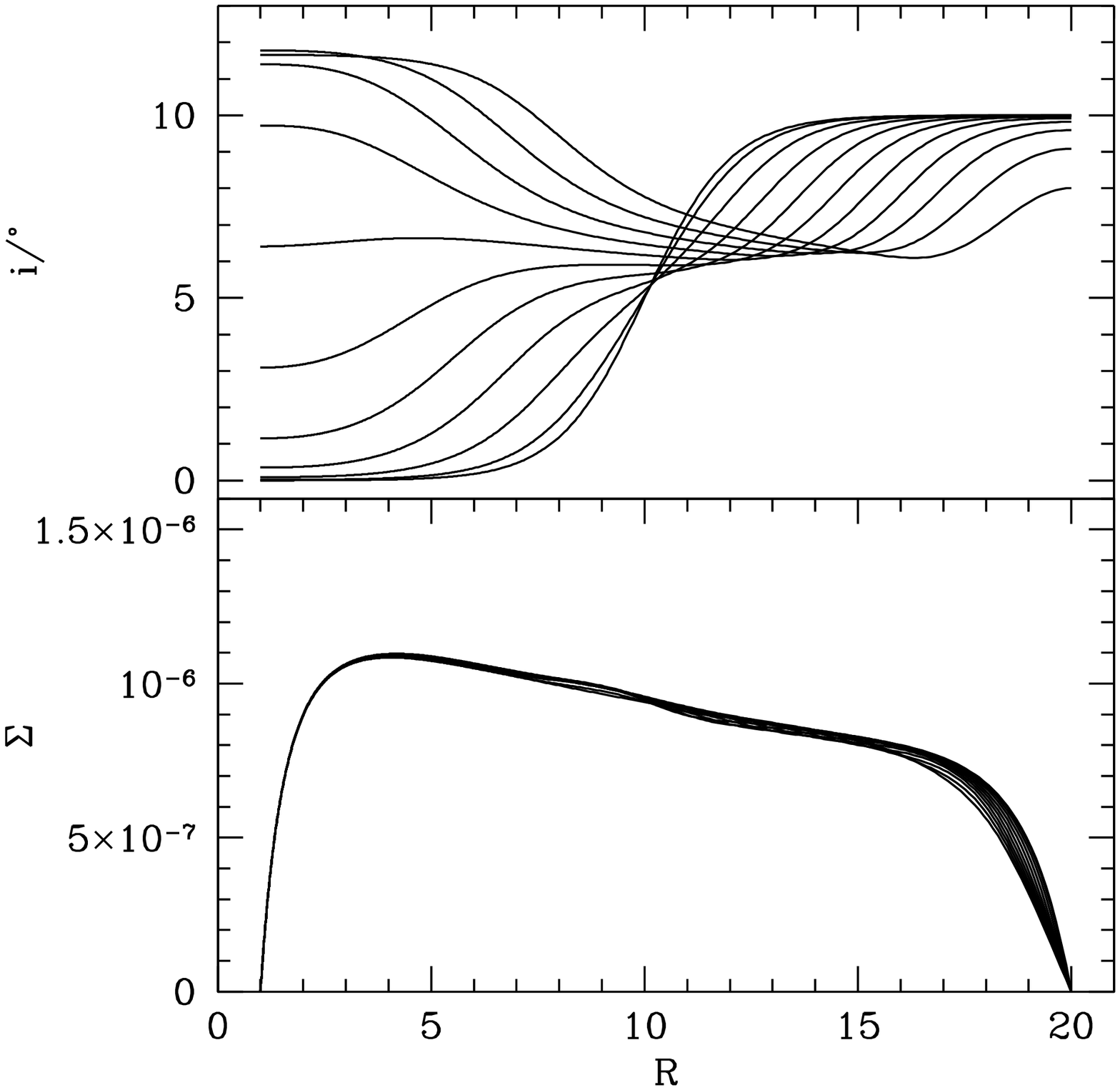}
\includegraphics[width=7.5cm]{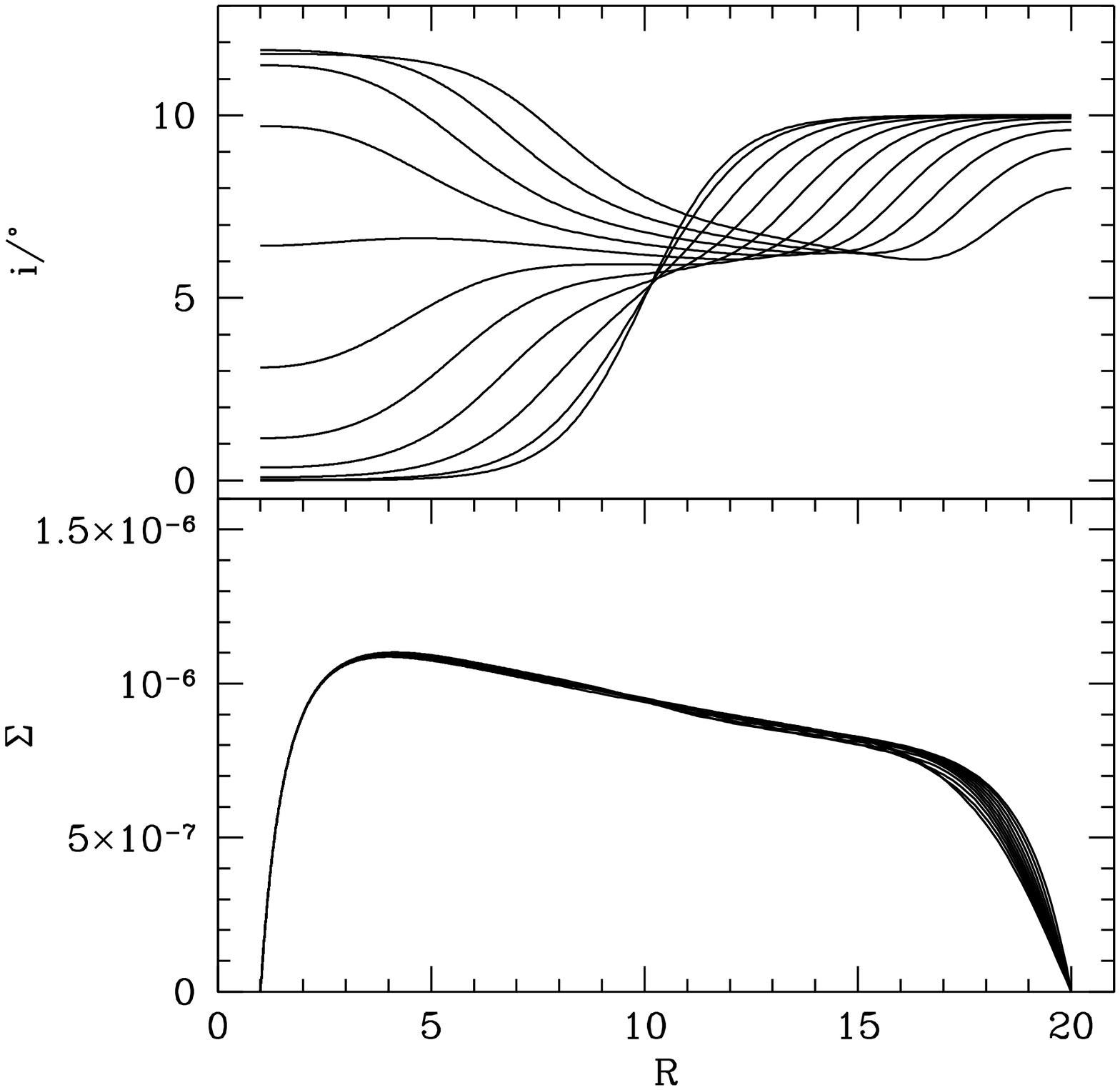}
\caption{Evolution of an initially warped disk around a single object
  with no external torque with $\alpha=0.01$ and $H/R=0.1$ (in the
  wave--like regime). The upper panels show the inclination and the
  lower panels the surface density.  Top left: The wave--like
  equations~(\ref{lo1}) and~(\ref{lo2}). The other panels solve the full equations~(\ref{main}) and~(\ref{flux}) with $\beta=0$ (top right), $\beta=1$ (bottom left) and $\beta=10$ (bottom right).  The times shown are every
  $10\,P_{\rm in}$.}
\label{wavelike}
\end{figure*}

We take the dot product of equation~(\ref{angmom}) with $\bm{l}$ and subtract $R^2 \Omega$ times equation~(\ref{mass}) to obtain an equation for the radial velocity
\begin{equation}
v_R=\frac{ \partial \bm{G}/\partial R \cdot \bm{l}}{R\Sigma \, d (R^2 \Omega)/ d R}.
\label{vr}
\end{equation}
We substitute the radial velocity equation~(\ref{vr}) into the conservation of mass equation~(\ref{mass}) to obtain an equation for the surface density evolution
\begin{align}
\frac{\partial \Sigma }{\partial t}  = 
- \frac{1}{R} \frac{\partial }{\partial R}\left[ \frac{\partial \bm{G}/\partial R \cdot \bm{l} }{ d (R^2 \Omega)/ d R}\right].
\end{align}
Further, we substitute the radial velocity equation~(\ref{vr}) into the conservation of angular momentum equation~(\ref{angmom}) to obtain an equation for the evolution of the angular momentum in the disk
\begin{align}
\frac{\partial \bm{L}}{\partial t}  =
& -\frac{1}{R} \frac{\partial }{\partial R}\left[ \frac{(\partial \bm{G}/\partial R \cdot \bm{l} )}{\Sigma \, d (R^2 \Omega)/d R}\bm{L}\right] 
+\frac{1}{R}\frac{\partial \bm{G}}{\partial R} +\bm{T}.
\end{align}

Since the disk is in  near--Keplerian rotation  we take
$\Omega\propto R^{-3/2}$ and find
\begin{align}
\frac{\partial \Sigma}{\partial t}= -\frac{2}{R}\frac{\partial }{\partial R}\left[\frac{(\partial \bm{G}/\partial R \cdot \bm{l} )}{R\Omega}\right]
\label{densevol}
\end{align}
and
\begin{align}
  \frac{\partial \bm{L} }{\partial t} =
  -\frac{2}{R}\frac{\partial}{\partial R} \left[ \left( 
\frac{( \partial \bm{G}/\partial R \cdot \bm{l})}{ \Sigma R\Omega} \right)\bm{L} \right]  +
\frac{1}{R}\frac{\partial \bm{G}}{\partial R}+\bm{T}.
%+ \bm{T}_{\rm p},
\label{main}
\end{align}

The key step to generalising the equations so that they are valid in
both diffusive and wave--like regimes is to now amend
equation~(\ref{lo1}) to read
\begin{align}
\frac{\partial \bm{G}}{\partial t}+  \omega \, \bm{l}\times \bm{G} &+
\alpha \Omega \bm{G}+\beta \Omega (\bm{G}\cdot \bm{l})\bm{l} = \cr
&\frac{\Sigma H^2
  R^3\Omega^3}{4}\frac{\partial \bm{l}}{\partial R} - \frac{3}{2}
(\alpha+\beta) \nu_1 \Sigma R^2 \Omega^2 \bm{l}.
\label{flux}
\end{align}
To effect this generalisation we have found it necessary to introduce
two extra terms dependent on a new dimensionless parameter
$\beta$. The fourth term on the left hand side has the effect of
damping the component of disk torque $\bm{G}$ perpendicular to the
local disk plane. The addition of the final term on the right hand
side is to add an additional shear viscosity term.  At this stage the
magnitude of $\beta$ is arbitrary, except that we shall require that
$\beta \gg \alpha$. We show the effects of different values for
$\beta$ in Section~\ref{numerical}.

\subsection{The new generalised equation in the two limits}

Equations~(\ref{main}), and~(\ref{flux}) provide a one-dimensional description of both the the disk surface density and the disk tilt. We now show that this generalised equation has the previous equations (Section~\ref{equations}) in both limits.
\begin{enumerate}
 \item In the wave-like limit we have $\alpha < H/R \ll 1$. The
   equations derived in this limit assumed that the surface density
   did not change with time, because in this limit the wave-like warp
   propagation happens on the shorter timescale than the viscous
   evolution of the surface density. Thus, in this limit, the final
   term on the right hand side of equation~(\ref{flux}) is
   negligible. In addition the assumption that $\Sigma$ is independent
   of time implies that $v_{\rm R} = 0$, unless there is an
     external source of mass. Thus (equation~\ref{vr}) we may take
   $\partial \bm{G}/\partial R \cdot \bm{l}=0$ and we may ignore the
   fourth term on the left hand side. Given this,
   equation~(\ref{flux}) now reduces to equation~(\ref{lo1}), as
   required. We note, however, that the full solution to the new
   equations allows for the evolution of the surface density also in
   the wave-like regime. The degree to which the surface density
   evolves depends on the magnitude of the new parameter $\beta$.

\item In the viscous limit $(\alpha > H/R)$, $\bm{G}$ evolves on a
  viscous timescale and so $\partial \bm{G}/\partial t \ll \alpha
  \Omega \bm{G}$ and we set $\partial \bm{G}/\partial t=0$.
  Furthermore, we set $\omega=0$ provided that $\omega \ll \alpha
  \Omega$ and we are left with
\begin{align}
\alpha \Omega \bm{G}+ &\beta \Omega (\bm{G}\cdot \bm{l})\bm{l} = \cr
&\frac{\Sigma H^2
  R^3\Omega^3}{4}\frac{\partial \bm{l}}{\partial R}  - \frac{3}{2}
(\alpha+\beta) \nu_1 \Sigma R^2 \Omega^2 \bm{l}.
\label{eq1}
\end{align}
We take the dot product of this with $\bm{l}$ to find an expression for 
$\bm{G}\cdot \bm{l} $ and then substitute that into equation~(\ref{eq1}) to find
\begin{equation}
\bm{G}=\frac{1}{2}\nu_2 \Sigma R^3 \Omega \frac{\partial \bm{l}}{\partial R} - \frac{3}{2}\nu_1 \Sigma R^2 \Omega \bm{l}.
\label{viscouslimit}
\end{equation}
Substituting this equation for $\bm{G}$ into equation~(\ref{main}), we recover the viscous disk evolution equation~(\ref{viscous}), which is valid for $H/r < \alpha \ll 1$ \citep{Ogilvie1999}. 

\end{enumerate}

\section{Numerical solutions}
\label{numerical}

We solve  equations~(\ref{main}) and~(\ref{flux}) as an initial
value problem for $\bm{L}$, and $\bm{G}$ using finite differences.
The method is first--order explicit in time. We use Cartesian
coordinates and treat each component of the vectors separately. The
units in the code are defined with $G=M=1$, where $M$ is the mass of
the central object. The Keplerian orbital period at the inner disk
radius $R_{\rm in}=1$ is $P_{\rm in}=2\pi$.  We take the boundary
conditions that $\bm{G}=\bm{0}$, $\Sigma=0$ and $\partial
\bm{l}/\partial R=0$ at $R=R_{\rm in}$ and $R=R_{\rm out}$.  The
initial condition on $\bm{G}$ is always taken as
$\bm{G}(R,0)=\bm{0}$.

We consider the evolution of an initially warped disk around a single
central object. There is no external torque on the disk, so $\bm{T}=0$
and $\omega=0$. The disk extends from $R_{\rm in}=1$ up to $R_{\rm
  out}=20$. We take the initial surface density of the disk to be
distributed as a simple power law with ends truncated at $R_{\rm in}$ and $R_{\rm out}$
\begin{equation}
\Sigma(R,0)=\Sigma_0 \left(\frac{R}{R_{\rm in}}\right)^{-1/2}\left[1-\left(\frac{R_{\rm in}}{R}\right)^{\frac{1}{2}}\right]\left[1-e^{R-R_{\rm out}}\right].
\end{equation}
The constant $\Sigma_0$ is arbitrary as the equations are linear in
$\Sigma$. Here we have scaled the total disk mass to be $0.001\,M$.
The first factor in brackets on the RHS is a power law that represents
a steady disk with $\nu_1 \Sigma \propto \,\rm const$ if mass is added
at the outer edge. The second and third factors enforce zero torque
($\Sigma=0$) inner and outer boundary conditions, respectively.  Note
that the surface density is not in steady state since we do not add
material to the disk.

The initial tilt of the disk is described by
\begin{equation}
i(R,0)=10^\circ \,\left[\frac{1}{2}\tanh \left( \frac{R-R_{\rm warp}}{R_{\rm width}}\right)+\frac{1}{2}\right].
\end{equation}
Since the equations are linear in disk tilt, the normalisation of $i$ is arbitrary.
The disk has an inclination of zero at the inner disk edge, an
inclination of $10^\circ$ at the outer disk edge and a warp at radius $R_{\rm
  warp}=10$ with a width of radius $R_{\rm width}=2$.  There is no twist in
the disk. Since we do not have any torques to cause precession, the
disk remains untwisted throughout its evolution. Thus we consider only the inclination of
the disk and not the nodal precession angle.

\subsection{Wave--like propagation; $\alpha< H/R$}
\label{wl}

We consider the evolution of an initially warped disk with parameters
in the wave--like limit, $\alpha=0.01$ and $H/R=0.1$.
Figure~\ref{wavelike} shows the disk inclination and surface density
evolution for several cases. In the top left panel we solve the
wave--like warped disk equations~(\ref{lo1}) and~(\ref{lo2}) with a
fixed density distribution. The warp in the disk propagates both
inwards and outwards. The inwards propagating wave reflects off the
inner boundary and then begins to propagate outwards.

In the other panels of Fig.~\ref{wavelike} we solve the full disk
equations~(\ref{main}) and~(\ref{flux}) with different values for
$\beta$.  In the top right panel we show the behaviour that occurs if
we do not introduce the parameter $\beta$. With $\beta=0$ the result
is that there appears to be unphysical evolution of the disk surface
density which occurs where the initial warp change was strongest, and
which continues long after the initial warp has propagated away.  
  The surface density anomaly shuld not keep growing at the position
  of the initial tilt change, even when the tilt at that point has
  evolved elsewhere.  Furthermore, this behaviour is not seen in three
  dimensional hydrodynamical simulations \citep[e.g.][]{Nealon2015}.
This unphysical behaviour was the reason for introducing the new
parameter $\beta$. The inclination evolution is very similar when we
solve the wave--like equations (top left panel) or the full equations
for $\beta \gtrsim 1$ (bottom panels). However, there is surface
density evolution when we solve the full equations and angular
momentum is conserved.  The bottom two panels show that for $\beta
\gtrsim 1$, the surface density evolution is independent of the value
for $\beta$.  There is slight difference between $\beta=1$ and
$\beta=10$ but we find no difference for even higher $\beta$ compared
to $\beta=10$.

\begin{figure}
\centering
\includegraphics[width=7.5cm]{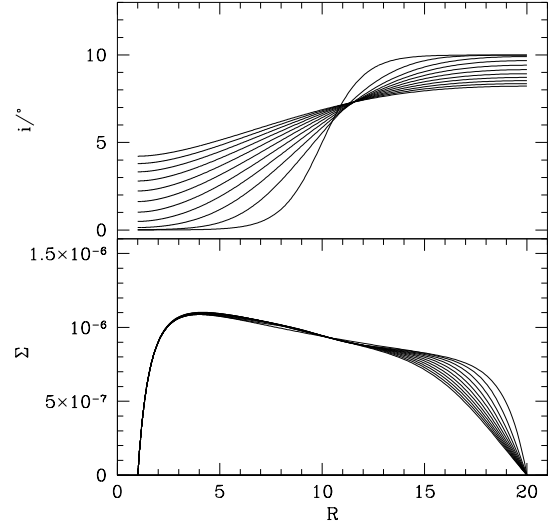}
\caption{Evolution of an initially warped disk around a single object
  with no external torque with $\alpha=0.1$ and $H/R=0.01$ (in the
  viscous regime). The upper panels show the inclination and the
  lower panels the surface density. 
  The full equations are solved with $\beta=10$.  The times shown are every
  $500\,P_{\rm in}$ and as time advances the inclination at the inner
  edge of the disk increases.}
\label{diffusive}
\end{figure}

\subsection{Diffusive warp propagation; $\alpha> H/R $}
\label{diff}

As a check, we consider the evolution of a disk with parameters in the
diffusive regime.  We take $\alpha=0.1$ and
$H/R=0.01$. Fig.~\ref{diffusive} shows the disk inclination and
surface density evolution solving the full equations~(\ref{main})
and~(\ref{flux}) with  $\beta=10$. We have also solved the diffusive equation~(\ref{viscous}) but find there is no  difference between the two solutions  and so we do not show this. In the diffusive regime there is no difference between solving the diffusive equations and the full equations that we have derived. The additional $\beta$ damping term has no effect in this limit.

\begin{figure}
\centering
\includegraphics[width=7.5cm]{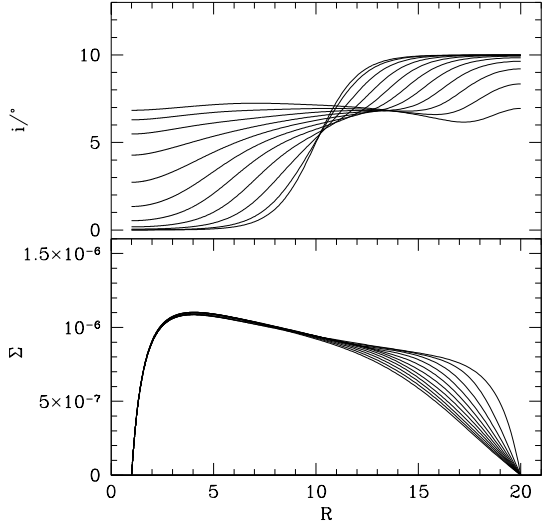}
\caption{Evolution of an initially warped disk around a single object
  with no external torque with $\alpha=0.1$ and $H/R=0.1$ (in the
 intermediate regime). The full equations are solved with $\beta=10$. The upper panels show the inclination and the
  lower panels the surface density.   The times shown are every
  $10\,P_{\rm in}$ and as time advances the inclination at the inner
  edge of the disk increases.}
\label{intermediate}
\end{figure}

\subsection{Intermediate regime; $\alpha=H/R$}

Neither the wave-like equations nor the diffusive equations are able to model the evolution of a disk with $\alpha \approx H/R$. However, the full equations we have developed, equations~(\ref{main})
and~(\ref{flux}), can be used in this regime. Fig.~\ref{intermediate} shows the solution to the full equations with $\beta=10$ for a disk with $\alpha=0.1$ and $H/R=0.1$. The inner parts of the disk appear more diffusive in nature and the outer parts look more wave-like in the inclination evolution.

\section{Conclusions}
\label{conc}

We have introduced a new set of equations that describe the evolution of disk warp and of disk surface density in both low viscosity and high viscosity disks. We have shown that the two sets of equations agree with the equations for warp propagation previously derived in the two distinct regimes of low viscosity (wave-like warp propagation) and of high viscosity (diffusive warp propagation). In order to achieve this we have introduced a new dimensionless parameter $\beta$ which has the dominant effect of preventing unphysical evolution of surface density in the wave-like regime. We have not been able to determine the required magnitude of $\beta$ except to note that for $\beta\gg \alpha$ the unphysical evolution of surface density in the wave-like regime no longer occurs. In order to determine the value of $\beta$, and indeed to determine whether or not the new equations we present here provide an adequate description of warp evolution in general, it will be necessary to undertake a detailed analytic analysis \citep[c.f.][]{Ogilvie1999} and/or compare with detailed numerical simulations.

\section*{Acknowledgments}

RGM, SHL and AF acknowledge support from NASA through grant
NNX17AB96G. RN has received funding from the European Research Council
(ERC) under the European Union’s Horizon 2020 research and innovation
programme (grant agreement No 681601). CJN is supported by the Science
and Technology Facilities Council (grant number ST/M005917/1).

\bibliographystyle{apj} 
%\bibliography{martin}

\begin{thebibliography}{32}
\expandafter\ifx\csname natexlab\endcsname\relax\def\natexlab#1{#1}\fi

\bibitem[{{Bardeen} \& {Petterson}(1975)}]{BP1975}
{Bardeen}, J.~M. \& {Petterson}, J.~A. 1975, ApJl, 195, L65+

\bibitem[{{Bate} {et~al.}(2000){Bate}, {Bonnell}, {Clarke}, {Lubow}, {Ogilvie},
  {Pringle}, \& {Tout}}]{Bateetal2000}
{Bate}, M.~R., {Bonnell}, I.~A., {Clarke}, C.~J., {Lubow}, S.~H., {Ogilvie},
  G.~I., {Pringle}, J.~E., \& {Tout}, C.~A. 2000, MNRAS, 317, 773

\bibitem[{{Facchini} {et~al.}(2013){Facchini}, {Lodato}, \&
  {Price}}]{Facchinietal2013}
{Facchini}, S., {Lodato}, G., \& {Price}, D.~J. 2013, MNRAS, 433, 2142

\bibitem[{{Herrnstein} {et~al.}(1996){Herrnstein}, {Greenhill}, \&
  {Moran}}]{Herrnstein1996}
{Herrnstein}, J.~R., {Greenhill}, L.~J., \& {Moran}, J.~M. 1996, \apjl, 468,
  L17

\bibitem[{{King} {et~al.}(2013){King}, {Livio}, {Lubow}, \&
  {Pringle}}]{Kingetal2013}
{King}, A.~R., {Livio}, M., {Lubow}, S.~H., \& {Pringle}, J.~E. 2013, MNRAS,
  431, 2655

\bibitem[{{Lai}(1999)}]{Lai1999}
{Lai}, D. 1999, \apj, 524, 1030

\bibitem[{{Larwood} {et~al.}(1996){Larwood}, {Nelson}, {Papaloizou}, \&
  {Terquem}}]{Larwoodetal1996}
{Larwood}, J.~D., {Nelson}, R.~P., {Papaloizou}, J.~C.~B., \& {Terquem}, C.
  1996, MNRAS, 282, 597

\bibitem[{{Lodato} \& {Facchini}(2013)}]{Lodato2013}
{Lodato}, G. \& {Facchini}, S. 2013, \mnras, 433, 2157

\bibitem[{{Lubow} \& {Ogilvie}(2000)}]{LO2000}
{Lubow}, S.~H. \& {Ogilvie}, G.~I. 2000, ApJ, 538, 326

\bibitem[{{Lubow} {et~al.}(2002){Lubow}, {Ogilvie}, \& {Pringle}}]{LO2002}
{Lubow}, S.~H., {Ogilvie}, G.~I., \& {Pringle}, J.~E. 2002, \mnras, 337, 706

\bibitem[{{Martin}(2008)}]{Martin2008}
{Martin}, R.~G. 2008, MNRAS, 387, 830

\bibitem[{{Martin} \& {Lubow}(2017)}]{Martin2017}
{Martin}, R.~G. \& {Lubow}, S.~H. 2017, \apjl, 835, L28

\bibitem[{{Martin} {et~al.}(2007){Martin}, {Pringle}, \&
  {Tout}}]{Martinetal2007}
{Martin}, R.~G., {Pringle}, J.~E., \& {Tout}, C.~A. 2007, MNRAS, 381, 1617

\bibitem[{{Martin} {et~al.}(2009){Martin}, {Pringle}, \&
  {Tout}}]{Martinetal2009}
---. 2009, MNRAS, 400, 383

\bibitem[{{Martin} {et~al.}(2011){Martin}, {Pringle}, {Tout}, \&
  {Lubow}}]{Martinetal2011}
{Martin}, R.~G., {Pringle}, J.~E., {Tout}, C.~A., \& {Lubow}, S.~H. 2011,
  MNRAS, 416, 2827

\bibitem[{{Nealon} {et~al.}(2018){Nealon}, {Dipierro}, {Alexander}, {Martin},
  \& {Nixon}}]{Nealon2018}
{Nealon}, R., {Dipierro}, G., {Alexander}, R., {Martin}, R.~G., \& {Nixon}, C.
  2018, \mnras, 481, 20

\bibitem[{{Nealon} {et~al.}(2015){Nealon}, {Price}, \& {Nixon}}]{Nealon2015}
{Nealon}, R., {Price}, D.~J., \& {Nixon}, C.~J. 2015, \mnras, 448, 1526

\bibitem[{{Nixon} \& {King}(2016)}]{Nixon2016}
{Nixon}, C. \& {King}, A. 2016, in Lecture Notes in Physics, Berlin Springer
  Verlag, Vol. 905, Lecture Notes in Physics, Berlin Springer Verlag, ed.
  F.~{Haardt}, V.~{Gorini}, U.~{Moschella}, A.~{Treves}, \& M.~{Colpi}, 45

\bibitem[{{Ogilvie}(1999)}]{Ogilvie1999}
{Ogilvie}, G.~I. 1999, MNRAS, 304, 557

\bibitem[{{Ogilvie}(2001)}]{Ogilvie2001}
---. 2001, MNRAS, 325, 231

\bibitem[{{Ogilvie}(2006)}]{Ogilvie2006}
---. 2006, MNRAS, 365, 977

\bibitem[{{Ogilvie} \& {Dubus}(2001)}]{OD2001}
{Ogilvie}, G.~I. \& {Dubus}, G. 2001, MNRAS, 320, 485

\bibitem[{{Papaloizou} \& {Lin}(1995)}]{PL1995}
{Papaloizou}, J.~C.~B. \& {Lin}, D.~N.~C. 1995, ApJ, 438, 841

\bibitem[{{Papaloizou} \& {Pringle}(1983)}]{PP1983}
{Papaloizou}, J.~C.~B. \& {Pringle}, J.~E. 1983, MNRAS, 202, 1181

\bibitem[{{Papaloizou} \& {Terquem}(1995)}]{PT1995}
{Papaloizou}, J.~C.~B. \& {Terquem}, C. 1995, MNRAS, 274, 987

\bibitem[{{Pringle}(1981)}]{Pringle1981}
{Pringle}, J.~E. 1981, ARA\&A, 19, 137

\bibitem[{{Pringle}(1992)}]{Pringle1992}
---. 1992, MNRAS, 258, 811

\bibitem[{{Pringle}(1996)}]{Pringle1996}
---. 1996, MNRAS, 281, 357

\bibitem[{{Pringle}(1999)}]{Pringle1999}
{Pringle}, J.~E. 1999, in Astronomical Society of the Pacific Conference
  Series, Vol. 160, Astrophysical Discs - an EC Summer School, ed.
  {J.~A.~Sellwood \& J.~Goodman}, 53

\bibitem[{{Scheuer} \& {Feiler}(1996)}]{SF1996}
{Scheuer}, P.~A.~G. \& {Feiler}, R. 1996, MNRAS, 282, 291

\bibitem[{{Shakura} \& {Sunyaev}(1973)}]{SS1973}
{Shakura}, N.~I. \& {Sunyaev}, R.~A. 1973, A\&A, 24, 337

\bibitem[{{Wijers} \& {Pringle}(1999)}]{WP1999}
{Wijers}, R.~A.~M.~J. \& {Pringle}, J.~E. 1999, MNRAS, 308, 207

\end{thebibliography}

\label{lastpage}
\end{document}